\documentclass[10pt,conference]{IEEEtran}
\IEEEoverridecommandlockouts
\usepackage[T1]{fontenc}
\usepackage{amsmath,amssymb}
\usepackage{booktabs}
\usepackage{array}
\usepackage{graphicx}
\usepackage{cite}
\usepackage{url}
\usepackage{xurl} %
\usepackage{tikz}
\usepackage[table]{xcolor}
\usepackage{hyperref} 
\hypersetup{pdfencoding=auto,
  pdftitle={When Instruction Files Go Stale: Documentation Debt and Hidden Cost in Coding Agents},
  pdfauthor={Tural Mehtiyev, Biruk Tadesse, Lawrence Arkoh, Wesley Assun\c{c}\~{a}o},
  pdfsubject={Software engineering; technical debt; LLM-based coding agents},
  pdfkeywords={documentation debt, technical debt, coding agents, LLM agents, repository instructions, AGENTS.md, CLAUDE.md, context files, stale documentation, agent trajectories, software maintenance, controlled evaluation}}
\usepackage[most]{tcolorbox}
\usepackage{balance} 

\newtcolorbox{rqanswer}{
    colframe=black,
    boxrule=0.7pt,
    arc=4pt,
    boxsep=2pt,
    left=3pt,
    right=3pt,
    top=3pt,
    bottom=3pt,
    before skip=4pt,
    after skip=4pt
}

\usetikzlibrary{positioning,arrows.meta}

\newcommand{\exRuns}{732}

\newcommand{\exTasks}{22}

\newcommand{\exBothOneStaleDead}{78}
\newcommand{\exBothOneStaleN}{80}

\newcommand{\stHaikuPBehavior}{$2\times10^{-28}$}

\newcommand{\stSonnetPBehavior}{$2\times10^{-33}$}

\begin{document}
\title{When Instruction Files Go Stale: Documentation Debt and Hidden Cost in Coding Agents\thanks{Preprint.}}
\author{\IEEEauthorblockN{Tural Mehtiyev, Biruk Tadesse, Lawrence Arkoh, Wesley Assun\c{c}\~{a}o}
\IEEEauthorblockA{North Carolina State University\\
\{tmehtiy, batadess, larkoh, wguezas\}@ncsu.edu}}
\maketitle
\begin{abstract}
LLM-based coding agents such as Claude Code are guided by repository instruction files (also called context files or agent configuration files) like AGENTS.md and CLAUDE.md, which are read in every agent session. As any documentation artifact, these files can fall behind the code and become stale (e.g., a test folder moves, but the file still points to the old one). In technical debt terms, such a stale instruction is documentation debt, and its interest is the cost it adds to every later agent session (e.g., additional tool calls to correct stale instruction). Evaluations that check only task success can miss this cost and the stale file can persist in the repository. In this study, we ask what leads agents to act on a stale instruction, how it affects their behavior and at what cost, and how they recover. We study three real-world stale instructions from active open-source repositories in depth, each a different kind of guidance: a test command, a code location, and a procedure for adding a component. Agents fix \exTasks{} maintenance tasks under three versions of each instruction (left stale, corrected to the current path, and removed), with the code and the task held fixed, and we analyze \exRuns{} trajectories of two models at the level of individual tool calls. We find that agents follow stale guidance when it is relevant to their tasks, sometimes even when correct information is available elsewhere. Following stale instructions can lead to wasted tool calls and increased effort without affecting task success. On the stale test command, Claude Sonnet 5.5 needed about two extra tool calls per run, a third more than with the corrected command; in the other cases, we could not detect a difference. Although agents generally recovered from the immediate failures, none repaired the underlying instruction file. Correcting the instructions eliminated operations on obsolete paths, while removing them prevented most. Our findings show that task success alone can hide the costs of documentation debt, highlighting the need to evaluate agent behavior beyond final outcomes.

\end{abstract}


\section{Introduction}\label{sec:intro}
\definecolor{trexp}{RGB}{226,234,246}
\definecolor{tredit}{RGB}{252,232,205}
\definecolor{trpass}{RGB}{214,240,214}
\definecolor{trdead}{RGB}{248,212,212}
\definecolor{trerr}{RGB}{232,232,232}
\newcommand{\trhead}[3]{\multicolumn{3}{@{}c@{}}{\textbf{#1}}\\[-1pt]\multicolumn{3}{@{}c@{}}{\scriptsize #2 \textbar{} #3 \textbar{} \textbf{bug fixed}}\\\midrule}
\newcommand{\trrow}[4]{\rowcolor{#1}#2 & #3 & #4\\}
\newcommand{\cmd}[1]{{\fontsize{6.3}{7.4}\selectfont\ttfamily #1}}
\newcommand{\trphase}[1]{\rowcolor{white}\multicolumn{3}{@{}l@{}}{\tiny\bfseries\MakeUppercase{#1}}\\}
\begin{figure*}[t]
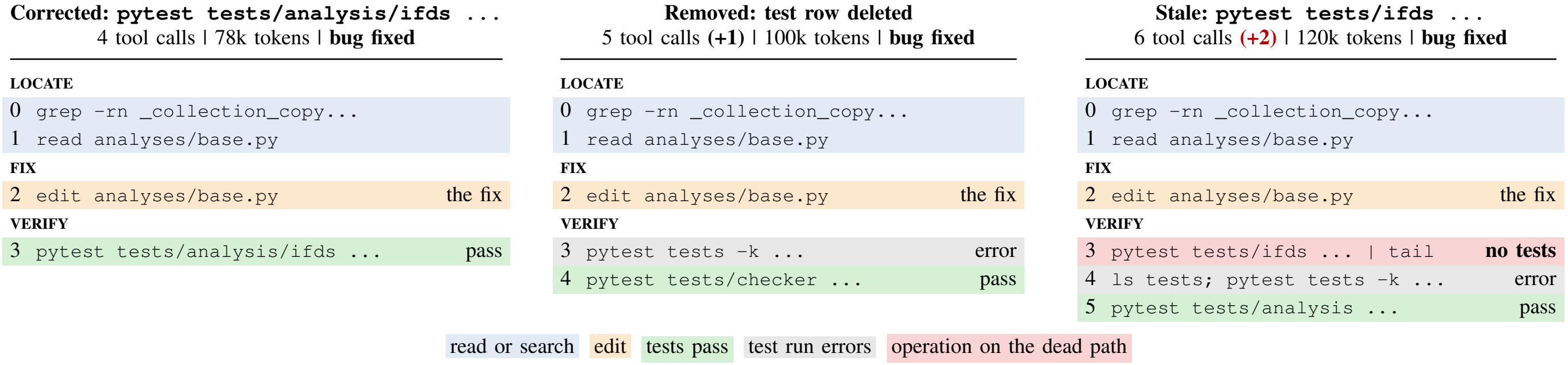

\centering
\scriptsize
\setlength{\tabcolsep}{2.5pt}
\renewcommand{\arraystretch}{1.15}
\begin{minipage}[t]{0.325\textwidth}\centering
\begin{tabular}[t]{@{}l >{\raggedright\arraybackslash}p{0.70\linewidth} r@{}}
\trhead{Corrected: \texttt{pytest tests/analysis/ifds \dots}}{4 tool calls}{78k tokens}
\trphase{Locate}
\trrow{trexp}{0}{\cmd{grep -rn \_collection\_copy\dots}}{}
\trrow{trexp}{1}{\cmd{read analyses/base.py}}{}
\trphase{Fix}
\trrow{tredit}{2}{\cmd{edit analyses/base.py}}{the fix}
\trphase{Verify}
\trrow{trpass}{3}{\cmd{pytest tests/analysis/ifds \dots}}{pass}
\end{tabular}
\end{minipage}\hfill
\begin{minipage}[t]{0.325\textwidth}\centering
\begin{tabular}[t]{@{}l >{\raggedright\arraybackslash}p{0.70\linewidth} r@{}}
\trhead{Removed: test row deleted}{5 tool calls \textbf{(+1)}}{100k tokens}
\trphase{Locate}
\trrow{trexp}{0}{\cmd{grep -rn \_collection\_copy\dots}}{}
\trrow{trexp}{1}{\cmd{read analyses/base.py}}{}
\trphase{Fix}
\trrow{tredit}{2}{\cmd{edit analyses/base.py}}{the fix}
\trphase{Verify}
\trrow{trerr}{3}{\cmd{pytest tests -k \dots}}{error}
\trrow{trpass}{4}{\cmd{pytest tests/checker \dots}}{pass}
\end{tabular}
\end{minipage}\hfill
\begin{minipage}[t]{0.325\textwidth}\centering
\begin{tabular}[t]{@{}l >{\raggedright\arraybackslash}p{0.70\linewidth} r@{}}
\trhead{Stale: \texttt{pytest tests/ifds \dots}}{6 tool calls \textcolor{red!70!black}{\textbf{(+2)}}}{120k tokens}
\trphase{Locate}
\trrow{trexp}{0}{\cmd{grep -rn \_collection\_copy\dots}}{}
\trrow{trexp}{1}{\cmd{read analyses/base.py}}{}
\trphase{Fix}
\trrow{tredit}{2}{\cmd{edit analyses/base.py}}{the fix}
\trphase{Verify}
\trrow{trdead}{3}{\cmd{pytest tests/ifds \dots{} | tail}}{\textbf{no tests}}
\trrow{trerr}{4}{\cmd{ls tests; pytest tests -k \dots}}{error}
\trrow{trpass}{5}{\cmd{pytest tests/analysis \dots}}{pass}
\end{tabular}
\end{minipage}

\vspace{4pt}
{\setlength{\fboxsep}{1.5pt}\colorbox{trexp}{read or search}\enspace\colorbox{tredit}{edit}\enspace\colorbox{trpass}{tests pass}\enspace\colorbox{trerr}{test run errors}\enspace\colorbox{trdead}{operation on the dead path}}

\caption{The stale test command in \texttt{pyflow}~\cite{repo-pyflow}, one of the three stale instructions we study: three runs of Claude Sonnet 5.5 on the same seeded bug (a negated condition), one per version of the line. Every tool call is shown (commands shortened), and (+$n$) counts calls beyond the corrected run. All three runs find and fix the bug the same way and differ only in checking the fix: one call with the corrected line, two without the line, and three with the stale line, whose documented command runs no tests while the pipe through \texttt{tail} reports success. All three pass our final check on the original tests. The two errors come from a test folder that needs a third-party module missing from the environment, unrelated to the bug. We chose the task whose runs varied least across repeats, and the repeat in which each run took its condition's median number of calls; the transcripts are in the replication package~\cite{replication2026}.}
\label{fig:trajectories}
\end{figure*}

Developers steer coding agents through instruction files such as \texttt{AGENTS.md} or \texttt{CLAUDE.md}. These files, usually stored at the root of a repository, tell the agent which commands build and test the project, where its parts live, and which instructions to follow~\cite{gloaguen2026agentsmd,lulla2026efficiency}. Agents read these files at the start of their sessions and follow their guidance closely~\cite{gloaguen2026agentsmd}. However, documentation tends to fall behind the code it describes: developers name outdated content among the documentation problems they encounter~\cite{aghajani2019docissues}, and practitioners report causes and effects of such documentation debt~\cite{rios2020documentation}. Instruction files are no exception. Some are never updated after they are created~\cite{mohsenimofidi2026context}, updates to others mostly add lines rather than remove them~\cite{chatlatanagulchai2025agentreadmes}, and Treude and Baltes flag potentially stale code references in the instruction files of 23.0\% of 356 repositories~\cite{treude2026contextrot}. Consequently, agents may follow outdated guidance, such as commands or paths that are no longer valid; in constructed conflicts between an instruction and the code, agents follow the instruction~\cite{xu2026stale,mohammadi2026workingset}.
We refer to an instruction that remains in the file after becoming outdated as a \emph{stale instruction}. 

Fig.~\ref{fig:trajectories} illustrates how such instruction can affect an agent's work. The example comes from the static-analysis framework \texttt{pyflow}~\cite{repo-pyflow}, whose \texttt{AGENTS.md} lists \texttt{pytest tests/ifds} as the command for running its IFDS tests. This command was valid until July 2026, when the tests moved to \path{tests/analysis/ifds}. However, the instruction remained unchanged through two subsequent edits of the file.
To examine its effect, we introduced a small bug by negating a condition, causing one test to fail. We then asked the same agent to fix the bug under three conditions: with the stale instruction, with the instruction corrected, and with it removed. In all three runs, the agent located and fixed the bug using the same sequence of actions: a search, a read, and an edit (see Fig.~\ref{fig:trajectories}). The difference appeared during verification. With the stale instruction, the agent ran the documented command, but the referenced directory no longer existed, so \texttt{pytest} executed no tests. Because the agent piped the output through \texttt{tail}, the shell reported a successful exit status. Although the output indicated that no tests had run, the agent needed two further test attempts before successfully verifying the fix. With the corrected instruction, verification took one call, while removing the instruction required two. All three runs fixed the bug and passed the final check on the original tests, making them indistinguishable to a success-only evaluation.

By observing this phenomenon through the lens of technical debt~\cite{kruchten2012technical}, the outdated or stale instruction is a documentation debt item, and the example shows both parts of that debt. The principal, the cost of repaying it, is a one-line edit. The interest is what each run pays while the line stays stale: here, two extra tool calls, half again the work of the corrected run, spent on a verification step whose failure looked like success. For one run this interest is small, but it is not paid once. The instruction file is loaded in every agent session, and the agent recovered without repairing the line, so every later task that reaches the line must pay again; if each pays a similar amount, the interest grows with the number of agent runs while the principal stays one line. Nothing in the outcome signals the debt, so it can stay: the bug is fixed, the tests pass, and the maintainer may learn nothing.

Existing research does not measure this kind of cost. Treude and Baltes detect stale references in instruction files and call for controlled experiments on how they affect agents, but they do not run agents themselves~\cite{treude2026contextrot}. Studies that do run agents compare runs with an instruction file against runs without one, and assume that the file's content is correct~\cite{gloaguen2026agentsmd,lulla2026efficiency}. Work on stale instructions builds conflicts over coding conventions, for example a file that demands one style while the existing code uses another, and judge the outcome by the code the agent produces~\cite{xu2026stale,mohammadi2026workingset}. In that setting, following the stale rule produces worse code, so checking the code reveals the harm. A stale path or command works differently: the repository itself contradicts it, so the agent hits an error, recovers, and still delivers a correct fix.

To investigate the consequences of stale instructions, our study is guided by three research questions (RQs):

\medskip
\noindent\textbf{RQ1.} What leads an agent to act on a stale instruction?

\smallskip
\noindent\textbf{RQ2.} How does a stale instruction affect the agent's behavior, and at what cost?

\smallskip
\noindent\textbf{RQ3.} How does the agent recover, and does it surface or repair the underlying stale instruction?
\par\medskip

To answer these questions, we study three real-world stale instructions in depth, each a different kind of guidance: a test command in \texttt{pyflow}~\cite{repo-pyflow}, a code location in \texttt{cog}~\cite{repo-cog}, and a procedure for adding a component in \texttt{aghub}~\cite{repo-aghub}. Each was written by the maintainers of an active repository, was correct until a documented code change made it outdated, and survived every later edit of its file (2, 13, and 4 edits; Fig.~\ref{fig:survival}). Our experiment asked the agents to fix the same bugs with the stale, corrected, and removed version of each instruction, on identical code and under the same setup: \exTasks{} tasks, three versions, five to seven repeats, and two models give \exRuns{} runs (Section~\ref{sec:study-design}). We measure task success, operations on the stale path, tool calls, and tokens, and rather than measuring many instructions shallowly, we follow every run tool call by tool call, which lets us explain why agents act as they do: how they came to take the stale path, what happened there, and how they recovered.

Agents took the stale path in two of the three cases: the test command in \exBothOneStaleDead{} of its \exBothOneStaleN{} stale runs, the procedure in about two thirds, and the code location never. The Location case is a deliberate contrast: its tasks work right beside the stale path but never need it. Whether agents used a stale line thus depended on the task more than on the line itself. Agents used it when the task required what the line names, such as running the tests or finding the code the procedure names, and offered no other starting point. They then followed it silently from the loaded file, sometimes even when the bug report named the correct location. At the stale path, the failure often looked like a success: the test command found no tests, but agents piped its output through \texttt{tail}, so the call was reported as successful, and in 19 of Sonnet's 40 stale runs the first check of the fix ran no tests. Per case, we detected extra effort only for Claude Sonnet 5.5 on the test command (about two tool calls per run, a third more than with the corrected line); in the other cases, we detected no difference in tool calls or tokens. Task success hid all of this, since agents fixed the bug in nearly every run under every version. Agents worked around the stale path within a few calls, mostly by searching the repository, told the user mainly when the failure hit the check of their own fix, and never repaired the instruction file, even when they noticed other outdated parts of it. Correcting the instruction removed the wasted work. Deleting it removed most of it but not all: agents still guessed the old folder, and as Fig.~\ref{fig:trajectories} shows, Sonnet needed as many tool calls without the test command as with the stale one, because it had to work out how to run the tests itself.

This paper makes three contributions. First, it compares stale, corrected, and removed versions of real-world stale instructions under controlled conditions and shows that a stale instruction can make agents waste work, including checks that test nothing, while task success stays the same. Second, it shows that this cost depends on whether the task leads the agent through the stale line, and that correcting a stale line works better than deleting it. Third, from the agents' trajectories, it explains why the debt stays hidden: agents meet the line where their workflow needs it, the failed check is reported as a success, and agents that notice a problem work around it without repairing the file. From this, we draw practical lessons: evaluations of instruction files should report what agents do, not only whether they succeed, and harnesses and agents can guard against stale paths and silent test failures.

\section{Related Work}\label{sec:related}

This section reviews documentation debt, repository instruction files and how they go stale, studies of whether instruction files help agents and of instructions that conflict with the code, and evaluations that look beyond task success. It then positions our study.

\noindent\textbf{Documentation debt.}
The technical-debt metaphor describes shortcuts that incur recurring costs, or \emph{interest}, until the \emph{principal} is repaid~\cite{cunningham1992wycash,kruchten2012technical}. Documentation that no longer matches the code is a recognized form of such debt~\cite{li2015mapping,alves2016identification}. Developers report outdated content among the documentation problems they encounter~\cite{aghajani2019docissues}, and Rios et al.\ survey practitioners about the causes, effects, and management of documentation debt~\cite{rios2020documentation}. Researchers have also developed techniques to detect documentation made obsolete by code changes, including comments inconsistent with source code~\cite{panthaplackel2021jit,liu2023obsolete} and outdated code-element references in README files and wikis~\cite{tan2024outdated}. For human readers, however, the effort associated with such debt can be difficult to observe: a developer may encounter an outdated reference, work around it, and leave no record of the detour. For coding agents, the same detour may appear in recorded tool calls, even when the task is completed successfully.

\noindent\textbf{Instruction files and how they go stale.}
Repository instruction files, such as \texttt{AGENTS.md} and \texttt{CLAUDE.md}, provide project-specific instructions to coding agents and are widely used to configure how they work~\cite{galster2026harness}. They commonly specify how to build and test a project, where code is located, and how to carry out routine changes~\cite{chatlatanagulchai2025agentreadmes,chatlatanagulchai2025manifests,mohsenimofidi2026context}. Previous studies report two maintenance patterns that may contribute to outdated instructions: some files are never updated after creation~\cite{mohsenimofidi2026context}, while updates to others consist mainly of small additions with few deletions~\cite{chatlatanagulchai2025agentreadmes}. Treude and Baltes describe inconsistencies between agent configuration files and repository code as \emph{context rot}~\cite{treude2026contextrot}. Applying a checker originally built to find outdated code references in README files~\cite{tan2024outdated} to the instruction files of 356 repositories, they flag potentially stale code-element references in 23.0\% of them and conclude that such inconsistencies are common. Their study focuses on detecting inconsistencies rather than observing agents. In their research roadmap, they call for controlled experiments to determine which forms of context rot affect coding-agent behavior. A \emph{stale instruction}, as we use the term, is one specific form of context rot: a line that was once correct and that a later code change contradicted. We use the narrower term because our experiment varies exactly one such line. Because agents use instruction files as they work, the consequences of an outdated reference may be visible in their recorded actions.

\noindent\textbf{Do instruction files help agents?}
Another line of research examines how instruction files affect coding agents, generally without testing whether their contents remain current. Gloaguen et al.\ compare agents given no instruction file, an LLM-generated file, or a developer-written file~\cite{gloaguen2026agentsmd}. They find no consistent improvement in task success, despite increased inference costs, and observe that agents follow the instructions they receive, for example by invoking tools named in the files. Lulla et al.\ evaluate an agent on 124 pull requests with and without \texttt{AGENTS.md}, reporting lower median runtime and output-token use when the file is present~\cite{lulla2026efficiency}. Shepard and Albrecht take a different approach, refining instruction files through synthetic bug-fix probes and finding that the refined instructions help agents locate relevant code and resolve more tasks~\cite{shepard2026probe}. Together, these studies show that instruction files can shape agent behavior, although their effects on success and efficiency vary across settings. They primarily examine whether instructions are provided or how to improve them, rather than isolating the consequences of instructions that were once correct but have become outdated.

\noindent\textbf{Stale and conflicting instructions.}
Two recent studies investigate constructed conflicts between repository instructions and existing code. Xu and Wu create 40 Python projects in which instruction files prescribe coding conventions that differ from those used in the code~\cite{xu2026stale}. Without intervention, Claude Code and Codex CLI follow the conflicting instructions in 90.1\% of scored choices, with Claude Code often acknowledging the conflict. Adding a sentence that directs agents to prioritize existing code reverses this tendency. Mohammadi et al., as part of a broader study of what agents need to know to edit coherently, compare three conditions: a standard that agrees with the code, no standard, and a standard that prescribes a less desirable practice~\cite{mohammadi2026workingset}. Agents choose the preferred implementation in 100\%, 33\%, and 0\% of scored decisions, respectively. In that setting, a conflicting standard leads to worse decisions than no standard. Both studies show that agents may follow incorrect rules about \emph{how code should be written}, and both assess the consequences through the resulting implementation decisions or code.

Our setting differs in the kind of outdated information and the consequences we examine. We study test commands, code locations, and development procedures that were once valid but became obsolete when the repository changed. Following such a reference need not damage the final repair: an agent may encounter the obsolete location, recover, and still complete the task correctly. The consequence may instead appear in the operations the agent performs along the way. We examine when agents use these references, what happens when they do, and how they recover.

\noindent\textbf{Looking beyond task success.}
Coding agents are often evaluated by whether their final patches pass benchmark tests. Several studies question whether passing tests adequately establish patch correctness: 29.6\% of plausible SWE-bench patches behave differently from the developers' fixes~\cite{wang2026solved}, while strengthened tests reject about one in five previously accepted patches from top agents~\cite{yu2026sweabs} and test augmentation reduces top agents' resolved rates by 4.2--9.0\%~\cite{li2026probe}. Other studies look beyond the final patch to examine the steps agents take. Analyses of agent trajectories associate context gathering before editing and subsequent validation with successful outcomes~\cite{bouzenia2025trajectories,mehtiyev2026beyond}, while another study finds that failures can begin early and remain undetected until recovery becomes difficult~\cite{zhao2026failure}. AgentLens identifies "Lucky Passes," in which agents pass benchmark tests despite problematic processes such as blind retries or missing verification~\cite{sahoo2026agentlens}. TraceProbe similarly shows that successful runs can differ in the amount of unsuccessful work they perform~\cite{shu2026traceprobe}.

These studies highlight two limitations of pass rates. First, passing tests may conceal an incorrect patch. Second, a successful repair may still involve ineffective or unnecessary work. Trajectory analyses can expose such behavior, but the observational studies discussed here do not isolate the effect of a particular outdated instruction. Moreover, a measure that categorizes a test invocation as validation may not distinguish a meaningful test run from one that executes no tests.

\noindent\textbf{Our position.}
We connect these lines of work. Prior work establishes that repository instructions can become outdated, that agents may follow information conflicting with the code, and that task success alone does not fully describe how agents work. What remains less understood is how a real-world stale instruction enters an agent's workflow, what happens when the agent acts on it, and whether recovery removes the source of the problem. We examine these questions by varying real-world stale references in repository instruction files. For each case, we compare three maintenance decisions: leaving the reference unchanged, correcting it, or removing it, while holding the code, task, model, and harness fixed. As in other controlled studies that vary one factor while keeping others constant~\cite{liu2024codeupdate,shi2026aletheia}, differences between conditions can be attributed to the instruction change within the studied tasks. Stale versus corrected evaluates the consequence of leaving an obsolete reference in place; corrected versus removed compares repairing the instruction with deleting it. In two of our three cases, obsolete references changed what agents did, most notably prompting a documented test command that ran no tests while appearing successful, without a detected difference in task success. Whereas constructed convention conflicts produced near-uniform instruction following~\cite{xu2026stale,mohammadi2026workingset}, the frequency of operations on obsolete references varied across our three cases, from nearly every run to none.

\section{Study Design}
\label{sec:study-design}
\newcommand{\stalepath}[1]{\textcolor{red!75!black}{#1}}
\newcommand{\entry}[4]{\node[ent] (#1) at (#2,#3) {\texttt{#4}};}
\newcommand{\gone}[4]{\node[ent, gonebox] (#1) at (#2,#3) {\texttt{\stalepath{#4}}};}
\newcommand{\note}[2]{\node[nt] at ([xshift=4pt]#1.east) {#2};}
\newcommand{\branch}[3]{\draw[br] (#1,#2) |- (#3.west);}
\begin{figure*}[t]
\centering
\scriptsize
\begin{tikzpicture}[
  ent/.style={anchor=west, inner sep=1pt, font=\scriptsize},
  gonebox/.style={draw=red!75!black, densely dashed, rounded corners=1pt, inner sep=1.5pt},
  nt/.style={anchor=west, inner sep=0pt, font=\scriptsize, text=black!65},
  br/.style={draw=black!45, line width=0.4pt},
  inst/.style={anchor=north west, draw=black!35, inner sep=3pt, align=left, font=\scriptsize, text width=5.05cm},
  ref/.style={-{Stealth[length=3pt]}, red!75!black, densely dashed, line width=0.5pt},
  ttl/.style={anchor=south west, inner sep=0pt, font=\small},
  tsk/.style={anchor=north west, inner sep=0pt, align=left, font=\scriptsize, text width=5.3cm}]
\def\r{0.40}\def\y{-1.45}
\begin{scope}[xshift=0cm]
\node[ttl] at (0,0.15) {\textbf{(a) Command case}\enspace\textcolor{black!60}{pyflow, Python}};
\node[inst] (i1) at (0,0) {\textcolor{black!60}{\texttt{AGENTS.md}}\\\texttt{pytest \stalepath{tests/ifds} tests/cli/...}};
\entry{c0}{0}{\y}{pyflow/}
\entry{c1}{0.35}{\y-\r}{src/pyflow/analysis/ifds/}
\entry{c2}{0.35}{\y-2*\r}{tests/}
\gone{c3}{0.70}{\y-3*\r}{ifds/}
\entry{c4}{0.70}{\y-4*\r}{analysis/ifds/}
\branch{0.12}{\y-0.15}{c1} \branch{0.12}{\y-0.15}{c2}
\branch{0.47}{\y-2*\r-0.15}{c3} \branch{0.47}{\y-2*\r-0.15}{c4}
\note{c1}{\textbf{8 bugs}}
\note{c3}{removed July 2026}
\note{c4}{moved here; checks every fix}
\draw[black!25] (0,\y-5.55*\r) -- (5.3,\y-5.55*\r);
\node[tsk] (tc) at (0,\y-5.75*\r) {\textcolor{black!60}{\textit{Tasks}}\enspace 8 bugs; 4 reports name the function.\\[1pt]\textbf{Every fix} is checked with the tests the stale command names.};
\end{scope}
\begin{scope}[xshift=6.0cm]
\node[ttl] at (0,0.15) {\textbf{(b) Location case}\enspace\textcolor{black!60}{cog, Python}};
\node[inst] (i2) at (0,0) {\textcolor{black!60}{\texttt{AGENTS.md}}\\\texttt{python/cog/\stalepath{base\_predictor.py}}: predictor interface};
\entry{l0}{0}{\y}{cog/}
\entry{l1}{0.35}{\y-\r}{python/cog/}
\gone{l2}{0.70}{\y-2*\r}{base\_predictor.py}
\entry{l3}{0.70}{\y-3*\r}{predictor.py}
\entry{l4}{0.70}{\y-4*\r}{\_inspector.py}
\branch{0.12}{\y-0.15}{l1}
\branch{0.47}{\y-\r-0.15}{l2} \branch{0.47}{\y-\r-0.15}{l3} \branch{0.47}{\y-\r-0.15}{l4}
\note{l2}{removed Feb.\ 2026}
\note{l3}{moved here; \textbf{1 bug}}
\note{l4}{\textbf{5 bugs}}
\draw[black!25] (0,\y-5.55*\r) -- (5.3,\y-5.55*\r);
\node[tsk] (tl) at (0,\y-5.75*\r) {\textcolor{black!60}{\textit{Tasks}}\enspace 6 bugs; 3 reports name the function.\\[1pt]\textbf{No task} needs the file the stale line names.};
\end{scope}
\begin{scope}[xshift=12.0cm]
\node[ttl] at (0,0.15) {\textbf{(c) Procedure case}\enspace\textcolor{black!60}{aghub, Rust}};
\node[inst] (i3) at (0,0) {\textcolor{black!60}{\texttt{CLAUDE.md}, to add an agent:}\\\texttt{1.\ \stalepath{crates/core/src/agents/}: create <name>.rs}};
\entry{p0}{0}{\y}{aghub/}
\entry{p1}{0.35}{\y-\r}{crates/}
\gone{p2}{0.70}{\y-2*\r}{core/src/agents/}
\entry{p3}{0.70}{\y-3*\r}{agents/src/}
\entry{p4}{1.05}{\y-4*\r}{agents/}
\entry{p5}{1.05}{\y-5*\r}{descriptor.rs}
\branch{0.12}{\y-0.15}{p1}
\branch{0.47}{\y-\r-0.15}{p2} \branch{0.47}{\y-\r-0.15}{p3}
\branch{0.82}{\y-3*\r-0.15}{p4} \branch{0.82}{\y-3*\r-0.15}{p5}
\note{p2}{removed March 2026}
\note{p4}{moved here; \textbf{4 bugs}}
\note{p5}{\textbf{4 bugs}}
\draw[black!25] (0,\y-5.55*\r) -- (5.3,\y-5.55*\r);
\node[tsk] (tp) at (0,\y-5.75*\r) {\textcolor{black!60}{\textit{Tasks}}\enspace 8 bugs; 4 reports name the function.\\[1pt]The line helps \textbf{when the agent must find} the descriptor code.};
\end{scope}
\end{tikzpicture}
\caption{The three studied cases and their tasks. Each instruction file names a path (red, dashed) that a later commit removed when the code moved; the tree shows where it moved and where the bugs of the case's tasks lie, and the block below summarizes the tasks. Instruction text quoted from the files and shortened; all \exTasks{} tasks are listed in the replication package~\cite{replication2026}.}
\label{fig:cases}
\end{figure*}
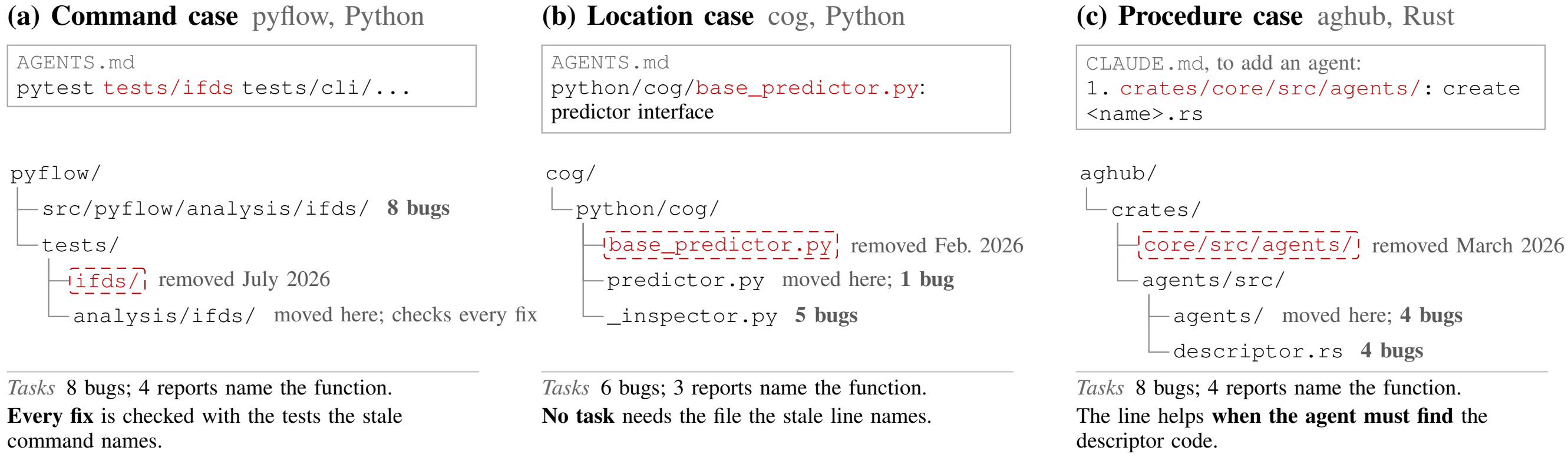

In this section, we describe the methodology used to investigate how coding agents interact with stale repository instructions. We begin by explaining how we selected three real-world stale instructions for in-depth study (Section~\ref{sec:case-selection}). We then introduce the selected cases and the kinds of guidance they provide  (Section~\ref{sec:cases}). Next, we describe the maintenance tasks constructed for each case (Section~\ref{sec:tasks}) and the experimental setup used to compare stale, corrected, and removed instructions (Section~\ref{sec:experimental-setup}). Across \exTasks{} maintenance tasks and two models, we hold the code and tasks fixed while measuring task success, operations on obsolete paths, tool calls, and token consumption. These measurements allow us to examine when agents follow stale instructions (RQ1), the costs they incur (RQ2), and how they recover, what they report, and whether they repair the underlying instruction files (RQ3).

\begin{figure}[t]
\centering
\includegraphics[width=\columnwidth]{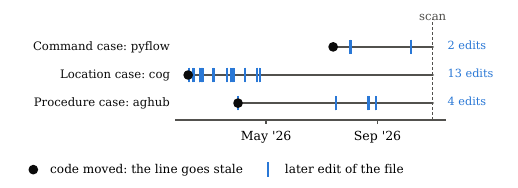}
\caption{Stale lines survive maintenance. Each row is the instruction file of one studied case, drawn from the day the code its stale line names moved (dot) to our scan (dotted line); ticks and the count on the right are later edits of the file. Every edit kept the stale line: 2 in the Command case (\texttt{pyflow}~\cite{repo-pyflow}), where the tests moved on July 14, 2026, 13 in the Location case (\texttt{cog}~\cite{repo-cog}), and 4 in the Procedure case (\texttt{aghub}~\cite{repo-aghub}).}
\label{fig:survival}
\end{figure}

\subsection{Case Selection}
\label{sec:case-selection}\label{sec:data-collection}
We searched GitHub code search for repositories with a root-level \texttt{AGENTS.md} or \texttt{CLAUDE.md} file. A case is one stale line in one of these repositories, and to answer our questions it had to meet five requirements. First, the repository is active: it is neither a fork nor archived, has at least five stars, and has at least one commit after the last change to its instruction file, so the code could have moved on since the file was written. Second, the instruction file names a path that existed in the repository until a code change moved it, while the file still presents the old path as current guidance; only real-world stale lines show how agents treat this debt as it occurs in practice. Third, the path moved rather than disappeared, so that a corrected version of the line exists; without one, we could not compare stale, corrected, and removed versions. Fourth, the line asks the agent to act, by running a command, going to a location, or following steps, so that whether the agent follows it can be observed. Fifth, the repository has a runnable test suite, so that tasks can be built from its tests and every run can be graded. The search queries, the sample, and the seed are in the replication package~\cite{replication2026}.

We kept only active repositories during the search and checked the other requirements in two steps. First, a script extracted every path the sampled files name (inline code and link targets) and checked each against the repository's current state and its commit history, flagging paths that once existed but no longer do; two coding-agent reviews (Claude and Codex) then confirmed whether each flagged line still presents its path as current guidance. Second, we checked the remaining requirements for each confirmed stale line by inspecting its repository and history.

Three stale lines met all selection criteria, one of each kind of instruction: a command to run, a location to go to, and a procedure to follow. We study all three in depth. In each, the line was correct before a documented change and wrong after it, with no alias keeping the old path working, and it stayed in its file through later edits, as Fig.~\ref{fig:survival} shows. The search served only to find cases, not to count them; it covers a small sample of repositories that use instruction files, and stale lines are common in practice~\cite{treude2026contextrot}. The extraction rules, review labels, and the outcome of each step are in the replication package~\cite{replication2026}.

\subsection{Studied Cases}\label{sec:cases}
Fig.~\ref{fig:cases} summarizes the three cases, which we name by the kind of guidance their stale line gives: the Command case, the Location case, and the Procedure case. Together they cover the three kinds of guidance that instruction files commonly give~\cite{chatlatanagulchai2025agentreadmes,mohsenimofidi2026context}: how to test the project, where code is located, and how to carry out a routine change. They also differ in how an agent uses them: a test command is executed as written, a code location is one of several ways to find code, and a procedure is followed step by step. This contrast in cases lets us examine when each kind of stale line enters an agent's work, what it costs, and how agents recover. Each case is described in detail in what follows:

\begin{itemize}
\item The \textbf{Command case} comes from \texttt{pyflow}~\cite{repo-pyflow}, a Python static-analysis framework. Its instruction file gives the command for running one group of tests. The tests moved to a new folder on July 14, 2026, but the command still names the old one, through two later edits of the file.
\item The \textbf{Location case} comes from \texttt{cog}~\cite{repo-cog}, a Python tool for packaging machine-learning models. Its instruction files name the file that holds the predictor interface, which has since moved to another file.
\item The \textbf{Procedure case} comes from \texttt{aghub}~\cite{repo-aghub}, a Rust project. Its instruction file lists the steps for adding an agent and places the new code in a folder that has since moved to a separate package; a second instruction file in the repository already gives the new location.
\end{itemize}

\subsection{Tasks}\label{sec:tasks}

Every task is a bug in the repository's own code that the repository's own tests catch. In the Command case, every fix is checked with the tests the stale command names; in the Procedure case, the agent needs the line when it must find the descriptor code. The Location case serves as a contrast. Its bugs lie in the same package as the file the stale line names, in the code that checks the predictor interface, but none of its tasks needs that file. The stale line is still in the agent's context in every session, and the agent works right beside the stale path. The case therefore tests whether a stale line pulls an agent off course when the work does not pass through it. Fig.~\ref{fig:cases} summarizes the tasks of each case; the full list is in the replication package~\cite{replication2026}.

\definecolor{armS}{RGB}{235,104,52}
\definecolor{armC}{RGB}{42,120,214}
\definecolor{armD}{RGB}{27,175,122}
\begin{figure*}[t]
\centering
\scriptsize
\begin{tikzpicture}[
  col/.style={font=\footnotesize\bfseries, align=center, text width=3.15cm},
  box/.style={draw=black!55, rounded corners=2pt, align=left, inner sep=3pt, text width=3.0cm, fill=white},
  stp/.style={draw=black!40, rounded corners=1.5pt, align=left, inner sep=2pt, text width=3.0cm, fill=black!3},
  arm/.style={draw=#1, rounded corners=2pt, align=left, inner sep=2.5pt, text width=3.0cm, fill=#1!12, line width=0.7pt},
  meas/.style={draw=black!45, rounded corners=1.5pt, align=left, inner sep=2pt, text width=3.0cm, fill=black!3},
  arr/.style={-{Stealth[length=4pt]}, black!65},
  node distance=0.18cm]
\node[col] (h1) at (0,0) {1. Real repository and task};
\node[col] (h2) at (3.55,0) {2. Build one workspace per run};
\node[col] (h3) at (7.1,0) {3. Only the stale line differs};
\node[col] (h4) at (10.65,0) {4. Fresh, isolated container};
\node[col] (h5) at (14.2,0) {5. Measure};
\node[box, below=0.25cm of h1] (src) {\textbf{Command case} (pyflow)\\\textbf{Location case} (cog)\\\textbf{Procedure case} (aghub)\\[3pt]
\exTasks{} tasks, each with one bug: a historical bug fix or a seeded bug in the code the line describes};
\node[stp, below=0.25cm of h2] (w1) {copy the repository at the task's commit, without its history};
\node[stp, below=of w1] (w2) {insert the task's bug};
\node[stp, below=of w2] (w3) {write the condition's instruction files};
\node[stp, below=of w3] (w4) {one fresh commit, so history does not reveal the bug};
\draw[arr] (w1) -- (w2); \draw[arr] (w2) -- (w3); \draw[arr] (w3) -- (w4);
\draw[arr] (src.east |- w1.west) -- (w1.west);
\node[arm=armS, below=0.25cm of h3] (S) {\textbf{Stale}: the line as found};
\node[arm=armC, below=of S] (C) {\textbf{Corrected}: names the location the code moved to};
\node[arm=armD, below=of C] (D) {\textbf{Removed}: line deleted, rest of the file kept};
\draw[arr] (w3.east) -- ++(0.15,0) |- (S.west);
\draw[arr] (w3.east) -- ++(0.15,0) |- (C.west);
\draw[arr] (w3.east) -- ++(0.15,0) |- (D.west);
\node[box, below=0.25cm of h4, fill=black!2, minimum height=2.55cm] (ctr) {};
\node[box, anchor=north, text width=2.85cm, fill=white] (agent) at ([yshift=-3pt]ctr.north) {\textbf{Claude Code}\\Claude Haiku 5.5 or Claude Sonnet 5.5};
\node[font=\scriptsize, align=left, text width=3.0cm, below=0.1cm of agent] (ch) {Instruction files loaded natively; bug report names no test path; no settings or memory from the host.};
\draw[arr] (S.east) -- (ctr.west |- S.east);
\draw[arr] (C.east) -- (ctr.west |- C.east);
\draw[arr] (D.east) -- (ctr.west |- D.east);
\node[meas, below=0.25cm of h5] (m1) {\textbf{Transcript}: operations on the stale path (primary); tool calls; tokens};
\node[meas, below=of m1] (m2) {\textbf{Final workspace}: restore the original tests, run the target tests and the suite offline $\rightarrow$ \textbf{bug fixed?}};
\draw[arr] (ctr.east |- m1.west) -- (m1.west);
\draw[arr] (ctr.east |- m2.west) -- (m2.west);
\end{tikzpicture}
\caption{How each run is built and graded. Every run starts from a real repository at the task's commit with one bug; between conditions, only the stale line of the instruction files differs. Each run is isolated in a fresh container, and outcomes are read from its transcript and from its final workspace, graded on the original tests.}
\label{fig:experiment}
\end{figure*}
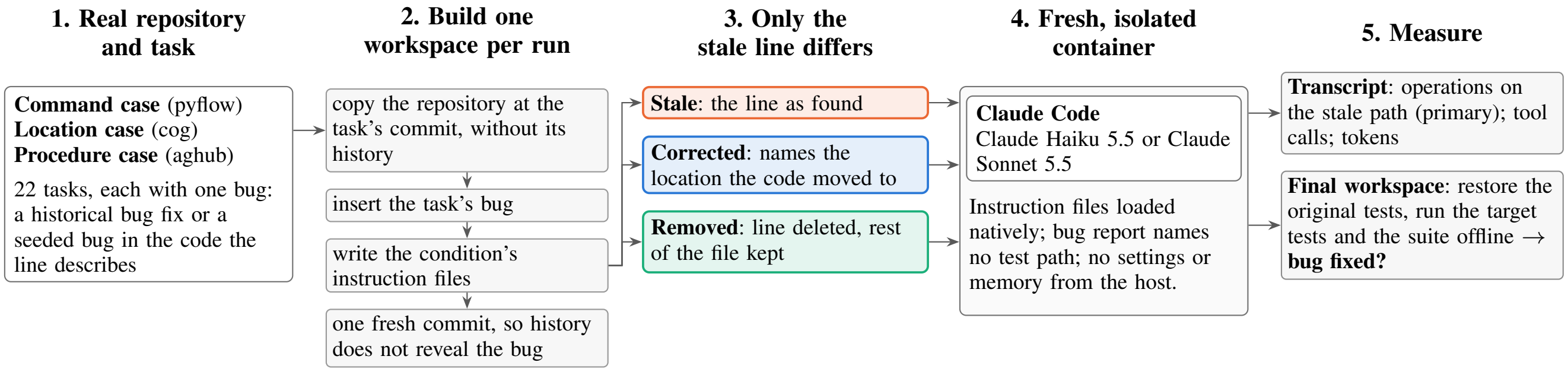

\subsection{Conditions, Runs, and Measures}
\label{sec:experimental-setup}
Fig.~\ref{fig:experiment} shows how each run is built and graded. Each task runs in three conditions, which correspond to the three actions a maintainer can take on the line: leave it stale; correct it to the location the code moved to, as recorded in the repository's history; or remove it. Each comparison isolates one question. Stale against corrected differs only in whether the path is right, so it measures what the wrong content itself costs, the interest of the debt, and the corrected runs show how the same task proceeds without it. Removed against the other two separates wrong guidance from missing guidance: it shows whether agents need the line at all, whether they return to the old path on their own, and whether deleting the line, the cheapest repair, is enough. Corrected and removed versions change every mention of the stale path in the instruction files; nothing else differs. 

We ran every task in each condition with two models, Claude Haiku 5.5 and Claude Sonnet 5.5, so that a finding does not depend on one model's habits; both ran in Claude Code, with about 40 runs per condition, case, and model (Table~\ref{tab:runs}).

\begin{table}[t]
\caption{Runs per case. Every task ran in all three conditions with both models.}
\label{tab:runs}
\centering
\scriptsize
\setlength{\tabcolsep}{4pt}
\begin{tabular}{@{}lccccc@{}}
\toprule
Case & Tasks & $\times$ Conditions & $\times$ Repeats & $\times$ Models & $=$ Runs\\
\midrule
Command case & 8 & 3 & 5 & 2 & 240\\
Location case & 6 & 3 & 7 & 2 & 252\\
Procedure case & 8 & 3 & 5 & 2 & 240\\
\midrule
Total & \exTasks{} & & & & \exRuns{}\\
\bottomrule
\end{tabular}
\end{table}

For each run we record whether the bug was fixed, whether the agent ran, read, or wrote the stale path (our primary measure), and its tool calls and tokens. To see why agents took the stale path and how they recovered, we coded every run that used it from its transcript with fixed rules: what the first stale call was for and what the agent did and said before it, where the call fell relative to the first edit, what it returned and whether the tool marked it as an error, how the agent reached the current location, and whether its messages named the stale path; for the test command, we also checked whether the first test run after the fix ran any tests. Setup details, the coding rules, and the per-run codes are in the replication package~\cite{replication2026}.

\noindent\textbf{Agent and runs.} The agent was Claude Code 2.1.316 at medium effort, with a 30-minute limit and a budget cap per run. Its prompt was the bug report alone, so it received the instruction files only through Claude Code's own loading, and it could run commands without asking for permission. Each block held one run of each condition for one task; the three runs ran together, and blocks ran in shuffled order, so the conditions shared time and machine load. Failed runs were repeated from scratch. A bug counted as fixed when the target tests and the suite passed offline on a copy of the final workspace, with separate test files restored.

\noindent\textbf{Measures and analysis.}
We measure token consumption across the entire run, including input, output, and cached tokens, accounting for context reused across model calls. We compare stale-path use and task success separately for each model using exact tests stratified by task, with Holm correction for multiple comparisons. We assess differences in effort using per-task medians and bootstrap confidence intervals. Our analyses of agent trajectories are exploratory and are used to explain how agents encounter stale instructions, respond to failures, and recover.

\section{Results}\label{sec:results}

\label{sec:rq4}

\begin{table*}[t]
\caption{Results per case and condition, for each model and for both together. Stale path: runs that ran, read, or wrote the stale path. Fixed: runs whose final workspace passed the original tests. Calls and tokens: per-task medians averaged over the case's tasks; for both models, the medians are over both models' runs of each task. In one test per model over all three cases, stale runs used the stale path more often than corrected runs, significantly after Holm correction (Haiku $p=$~\stHaikuPBehavior, Sonnet $p=$~\stSonnetPBehavior); the difference comes from the Command and Procedure cases, since no run in the Location case used the stale path.}
\label{tab:stage2}
\centering
\scriptsize
\setlength{\tabcolsep}{3pt}
\begin{tabular}{@{}llcccccccccccc@{}}
\toprule
 & & \multicolumn{4}{c}{Claude Haiku 5.5} & \multicolumn{4}{c}{Claude Sonnet 5.5} & \multicolumn{4}{c}{Both models}\\
\cmidrule(lr){3-6}\cmidrule(lr){7-10}\cmidrule(l){11-14}
Case & Condition & Stale path & Fixed & Calls & Tokens (k) & Stale path & Fixed & Calls & Tokens (k) & Stale path & Fixed & Calls & Tokens (k)\\
\midrule
Command case & Stale & 38/40 & 39/40 & 16.0 & 536 & 40/40 & 40/40 & 7.8 & 195 & 78/80 & 79/80 & 10.2 & 278\\
 & Corrected & 0/40 & 39/40 & 18.2 & 648 & 0/40 & 40/40 & 5.8 & 152 & 0/80 & 79/80 & 9.6 & 303\\
 & Removed & 0/40 & 39/40 & 14.0 & 454 & 1/40 & 40/40 & 7.8 & 222 & 1/80 & 79/80 & 10.9 & 352\\
\midrule
Location case & Stale & 0/42 & 42/42 & 7.5 & 222 & 0/42 & 42/42 & 5.3 & 157 & 0/84 & 84/84 & 6.1 & 175\\
 & Corrected & 0/42 & 42/42 & 7.3 & 220 & 0/42 & 42/42 & 5.5 & 162 & 0/84 & 84/84 & 6.3 & 186\\
 & Removed & 0/42 & 42/42 & 7.7 & 224 & 0/42 & 42/42 & 5.3 & 157 & 0/84 & 84/84 & 6.4 & 184\\
\midrule
Procedure case & Stale & 24/40 & 39/40 & 12.4 & 542 & 31/40 & 40/40 & 7.9 & 294 & 55/80 & 79/80 & 9.6 & 391\\
 & Corrected & 0/40 & 38/40 & 11.2 & 493 & 0/40 & 39/40 & 6.8 & 249 & 0/80 & 77/80 & 9.6 & 395\\
 & Removed & 13/40 & 39/40 & 10.5 & 402 & 7/40 & 40/40 & 7.8 & 283 & 20/80 & 79/80 & 8.9 & 339\\
\bottomrule
\end{tabular}

\end{table*}

Table~\ref{tab:stage2} gives the results for each case and condition. We report them in the order in which an agent meets a stale line: what leads it to take the stale path (RQ1), how the stale path affects its behavior and at what cost (RQ2), and how it recovers and whether it surfaces or repairs the stale line (RQ3).

\noindent\textbf{\emph{RQ1: What leads an agent to act on a stale instruction?}}\label{sec:selection}

\noindent\textbf{Agents took the stale path in two of the three cases.} With the stale line, agents used the stale path in 78 of 80 runs in the Command case, 55 of 80 in the Procedure case, and none of the 84 in the Location case (Table~\ref{tab:stage2}); with the corrected line, no run did. Stale against corrected is significant for each model after Holm correction (Haiku $p=$~\stHaikuPBehavior, Sonnet $p=$~\stSonnetPBehavior).

\noindent\textbf{The task decided whether the line was needed.} A stale line mattered only when the task led the agent to what the line describes. Every Command task ends with running the tests, so agents met the stale test command with either kind of bug report. The Procedure line says where the agent descriptors live, and the broad report pointed right there (``some agent descriptor behavior is wrong''), so both models used the stale folder in nearly every such run (Haiku 20, Sonnet 19 of 20); when the report named the function, Haiku used it in 4 of 20 runs and Sonnet in 12 of 20. In the Location case, the agent never had to look anything up: a report that named the function also named its file, for one bug the predictor file under its current name, and a broad report led the agent to list the package or run the tests, which showed the faulty file. With the right file in hand within two calls, no agent turned to the outdated location in the instruction file.

\noindent\textbf{The line filled a missing starting point.} With a broad report, Haiku's first call listed the stale folder (``I'll start by looking at the agent descriptors and the registry''), while Sonnet first ran the tests (18 of 19 runs) and then looked for the descriptor code of the failing test in the stale folder. With a named function, Haiku searched for it by name, but Sonnet searched inside the stale folder in 12 of 20 runs, although the report named the function's current module.

\noindent\textbf{Agents adopted the path silently.} Before their first stale call, agents cited the instruction file in 2 of the 133 runs that used the stale path and opened it in none; the path came from the file that Claude Code loads into the session. Their messages described the task, not where the path came from, and Sonnet copied the documented test command whole in 35 of its 40 runs in the Command case.

\noindent\textbf{Removing the line prevents most of it.} With the line removed, agents still used the stale path in 1 of 80 runs in the Command case and 20 of 80 in the Procedure case, all 20 with a broad report: no file named the old folder, but they guessed an \texttt{agents} folder inside the core crate from the code's structure. Only the corrected line prevented this.

\noindent\textbf{Answer to RQ1.} Agents took a stale path when the task led their strategy through what the line names and gave them no other starting point: always for the test command, mostly for the procedure when the report named only the area, and never for the code location. They took the path silently from the loaded file. Correcting the line prevented this in every run, and removing it prevented most of it.

\noindent\textbf{\emph{RQ2: How does a stale instruction affect the agent's behavior, and at what cost?}}\label{sec:encounter}

\noindent\textbf{The stale test command failed quietly.} Because the old test folder no longer exists, the documented command collected no tests and exited with an error. In 64 of the 80 stale runs in the Command case, at least one call on the stale path ran no tests, and Claude Code flagged none of these calls as an error: every run that ran the command piped its output through \texttt{tail} or \texttt{head}, so the shell reported success. In all 40 Sonnet runs, the first such call printed that the folder was not found and that no tests ran, yet it counted as a success; Haiku often hid the message itself (\texttt{2>/dev/null}, 16 of its 38 runs). As a result, 19 of Sonnet's 40 stale runs first checked their fix with a test run that ran no tests, against none of its 40 corrected runs. The missing folder of the Procedure case was flagged more often, in 20 of 24 Haiku runs and 13 of 31 Sonnet runs.

\noindent\textbf{When the agent met the line.} With a broad report, Sonnet met the stale command while reproducing the bug (20 of 20 runs); with a report that names the function, it met it while checking its fix (19 of 20). Haiku explored first whatever the report and met the command before editing in 34 of its 38 runs.

\begin{figure*}[t]
\centering
\includegraphics[width=\textwidth]{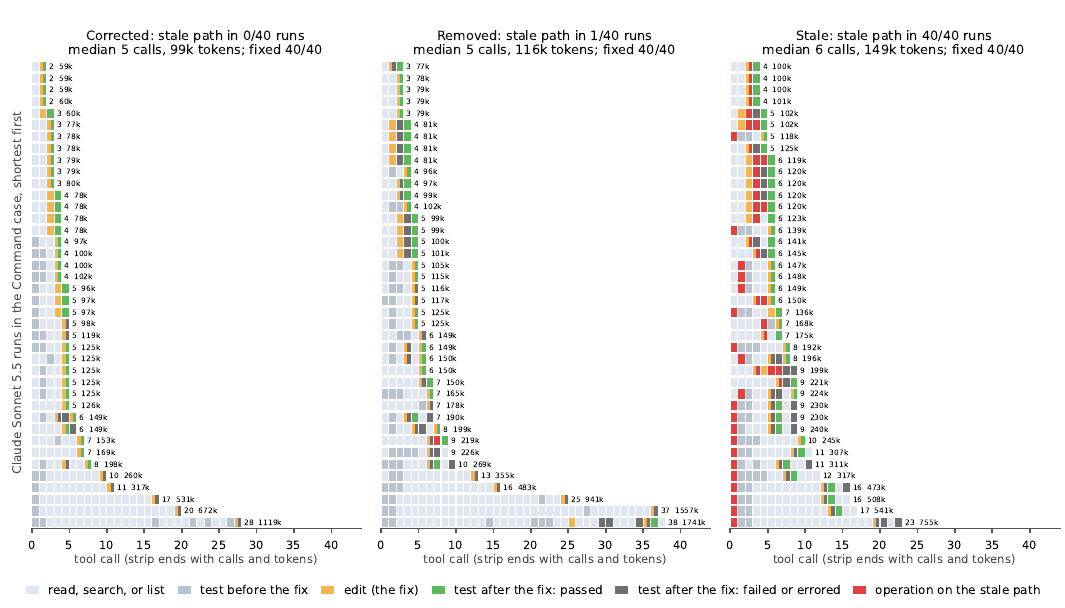}
\caption{The stale test command adds calls, mostly around and after the fix. Every run of Claude Sonnet 5.5 in the Command case, 40 per condition, as a strip of tool calls colored by what each call did; a half cell is one shell call that edits and then tests. Each strip ends with the run's tool calls and tokens. All 120 runs fixed the bug. The longest runs in every condition spend their calls reading and searching for the bug. In stale runs, the extra calls are operations on the stale path around the fix, usually the documented command, which runs no tests, followed by more test runs. Tokens grow with calls, but the per-case token difference is not resolved (Section~\ref{sec:rq4}).}
\label{fig:barcode}
\end{figure*}

\noindent\textbf{Agents still fixed the bug, but with more work.} Agents fixed the bug in nearly every run under every condition ($p=1.0$ for both models). For tool calls and tokens, we could resolve only one difference, and it is exploratory: on the test command, Sonnet's stale runs used 7.8 tool calls against 5.8 for its corrected runs, with the stale median higher in seven of the eight tasks; the 95\% interval excludes zero, though not after correcting for the number of estimates. Its tokens rose as well (195k against 152k), but not significantly. In four of these tasks, all 15 runs across the three conditions produced the same patch, byte for byte, so only the work to reach it differed (Fig.~\ref{fig:barcode}).

\noindent\textbf{The cost followed the report.} On the procedure, the stale line cost both models about two tool calls per task with a broad report (2.25 for Haiku, 2.0 for Sonnet) and almost none with a named function (0.0 and 0.25), even though Sonnet still entered the stale folder in most of those runs (RQ1). On the test command, Sonnet paid about two calls per task under both kinds of report, while reproducing the bug or while checking the fix. Haiku's runs varied too widely within a task to resolve a difference (on the test command, from 18 fewer to 9 more calls per task). All per-task comparisons are exploratory.

\noindent\textbf{Answer to RQ2.} A stale instruction changes what the agent does at the stale path: it runs a command or searches a folder that no longer exists, and the resulting failure often looks like a success, since a test command that runs no tests is reported as passing. The stale line cost a few tool calls each time an agent met it, measurably only for Sonnet on the test command (about two calls per run, a third more than with the corrected line; exploratory). Task success did not change, so an evaluation that checks only the outcome does not see the cost.

\noindent\textbf{\emph{RQ3: How does the agent recover, and does it surface or repair the underlying stale instruction?}}\label{sec:recovery}

\noindent\textbf{Most encounters ended in operational recovery.} Table~\ref{tab:rq3-recovery} starts from the 133 \emph{encounters}: stale-condition runs that operated on the studied obsolete path, which form the denominator for each row. The \emph{reached target} column records whether a later call actually ran tests off the stale path in the Command case or operated on the current folder in the Procedure case. By this measure, agents recovered in 130 of 133 encounters. The \emph{via search} column identifies the subset of recoveries preceded by a repository search; 98 recovered through search, most often Sonnet in the Procedure case (28 of 31), which searched the repository once the folder was missing, while 32 reached the target directly. Recovery took a median of two calls after the first stale-path operation; the case--model medians ranged from two calls for both models in the Command case to four for Haiku in the Procedure case. The other three trajectories did not reach the current target before they ended. These post-encounter sequences describe how agents worked around the stale guidance; because they include only runs that took the stale path, they are not causal estimates of its net cost.

\noindent\textbf{Recognition did not reliably reach the user.} The \emph{final disclosure} column is stricter than noticing a problem during execution: it counts a run only when the final response identifies the studied mismatch, not another documentation issue. Only 21 of the 133 encounters met this definition. Sonnet named the stale test folder in 17 of its 20 runs whose report named the broken function, where it usually met the command while checking its fix, but in none of its 20 runs with a broad report, where it met the command while reproducing the bug. Haiku named it in 3 of its 38 Command runs. In the Procedure case, only one final response reported the mismatch, describing the old layout without naming the folder, although Haiku often noted during the trajectory that the layout differed from \texttt{CLAUDE.md}; several Haiku reports instead discussed other outdated parts of the file. Thus, noticing a mismatch during execution did not imply that the maintainer received it in the final report. These report-level patterns are descriptive: report type is tied to the task rather than independently randomized.

\begin{table*}[!t]
\caption{Recovery and documentation-debt response after an agent operated on the stale path. Percentages use the encounters in the same row as the denominator. ``Via search'' is a subset of runs that reached the current target.}
\label{tab:rq3-recovery}
\centering
\small
\setlength{\tabcolsep}{5.5pt}
\begin{tabular}{llrccccc}
\toprule
& & & \multicolumn{2}{c}{Operational recovery, $n$ (\%)} & \multicolumn{3}{c}{Documentation-debt response, $n$ (\%)} \\
\cmidrule(lr){4-5}\cmidrule(lr){6-8}
Case & Model & Encounters & Reached target & Via search & Final disclosure & Suggested repair & Edited instruction \\
\midrule
Command   & Haiku  & 38 & 36 (95)  & 27 (71) & 3 (8)   & 0 (0) & 0 (0) \\
Command   & Sonnet & 40 & 40 (100) & 27 (68) & 17 (43) & 0 (0) & 0 (0) \\
\addlinespace[2pt]
Procedure & Haiku  & 24 & 23 (96)  & 16 (67) & 1 (4)   & 0 (0) & 0 (0) \\
Procedure & Sonnet & 31 & 31 (100) & 28 (90) & 0 (0)   & 0 (0) & 0 (0) \\
\bottomrule
\end{tabular}

\vspace{2pt}
\parbox{0.97\textwidth}{\footnotesize\emph{Note:} Final disclosure requires the final response to identify the studied mismatch, not another documentation issue. Suggested repair requires an explicit proposal to correct the instruction. No instruction file was edited in any of the 732 runs.}
\end{table*}

\noindent\textbf{Disclosure did not become repayment.} The last two columns distinguish recommending repayment from performing it. The \emph{suggested repair} column comes from exploratory coding and requires an explicit recommendation to correct the instruction; \emph{edited instruction} requires an actual change to the instruction file in the final workspace. Neither occurred. Final responses that identified the studied mismatch described the missing path or the replacement used to finish the task, but none proposed correcting the instruction, and no run in any condition edited an instruction file. The task asked the agent to fix a code defect, not to maintain documentation, so this is not evidence that editing the file was required for task success. It instead shows how the debt persists during ordinary maintenance: the agent can work around the wrong instruction, deliver a passing patch, and leave the same obstacle for the next session.

\noindent\textbf{Answer to RQ3.} Agents usually restored task progress within a few calls, often by searching for the current target, but recovery rarely produced an actionable maintenance signal. Mismatches noticed during a trajectory were often absent from the final report; reports that did name the studied mismatch did not recommend correcting it; and no run repaired the instruction file. Operational recovery therefore removed the immediate blockage without repaying the documentation debt.

\section{Discussion}\label{sec:discussion}

\noindent\textbf{The debt is paid in behavior, not in outcomes.} Agents fixed the bug in nearly every run, whatever the condition, but a stale line changed how they got there: they ran a test command that tested nothing, which was even reported as a success, or searched a folder that no longer exists. In technical debt terms, the principal is a one-line edit, and the interest is the extra work an agent does each time it meets the line. How much interest a line charges depends on what one encounter costs and on how often tasks reach the line; we measured only the first. It varied with the line and the model, and it was small when the task offered another way forward: with a report that named the function, even entering the stale folder cost Sonnet almost nothing. The debt can build up because task success, the measure most evaluations use, does not show it, and because later edits of the file left the stale line in place in all three cases (Fig.~\ref{fig:survival}).

\noindent\textbf{Not every stale line carries interest.} Agents followed the stale test command almost always, the stale procedure in about two thirds of runs, and the stale code location never. All three lines ask the agent to act; what separated them was whether the task led the agent through what the line names. Every fix ends with running the tests, and agents ran the documented command as written (Sonnet copied it whole in 35 of its 40 runs). The procedure mattered when a report gave no other starting point. No task needed the code location, so that case shows that this line did not matter for these tasks, not that code locations never matter. A stale line can also win over better information in the task: Sonnet searched the stale folder in 12 of its 20 runs even though the report named the function's correct module. Agents almost never said where the path came from (2 of 133 runs), so someone reading a trajectory cannot tell that the instruction file misled the agent. We therefore expect the interest to be largest for lines on a step most tasks share, such as the command that checks every fix, and maintainers with limited time may want to check those commands first. With three cases, the kind of instruction cannot be separated from its repository and tasks, so this remains a hypothesis for larger studies.

\noindent\textbf{Recovery hides the debt from those who could repay it.} The agent pays the interest, but only the maintainer can repay the principal, and recovery keeps the two apart. Agents reached the current location in 130 of the 133 runs that took the stale path, mostly by searching the repository (98 runs) rather than by correcting the path, so they finished their task without fixing the file. Whether the maintainer heard about the problem depended on when the agent met it. Sonnet mentioned the stale test folder in its final report in 17 of its 20 runs with a report that named the function, where it usually met the command while checking its fix, but in none of its 20 runs with a broad report, where it met the command while reproducing the bug. In the Procedure case, only one final report described the old layout. Agents also noticed other outdated parts of \texttt{CLAUDE.md} and left them: in 22 of Haiku's 120 runs in the Procedure case, under all three conditions, the final report pointed them out and often said the file was left alone because it was not part of the bug. No report proposed correcting the stale line, and no run edited an instruction file. An agent's workaround leaves a record only in a transcript that maintainers rarely read. Our tasks asked for a code fix, not documentation upkeep, so this is what agents do by default, not what they would do if asked.

\noindent\textbf{Models pay the debt differently.} Both models followed the stale lines at similar rates, and both went to the stale Procedure folder in nearly every run whose report named only the area. They split when the report named the function: Haiku then searched by name and used the folder in 4 of 20 runs, Sonnet in 12. The cost also showed differently. Sonnet's runs were short and consistent, so its extra tool calls on the test command could be measured; Haiku's runs varied so much that no difference in effort was detectable, although it followed the stale lines just as often. A cost measured with one model therefore says little about another, so evaluations of instruction files should use more than one model and report what agents did, not only how much effort they spent.

\noindent\textbf{For maintainers: correct the line rather than delete it.} Deleting a stale line removed most of its effect, but only correcting it removed all of it. With the line deleted, agents used the stale path in 1 of 80 runs in the Command case and 20 of 80 in the Procedure case, against 78 and 55 with the stale line and none with the corrected line; in the Procedure case, they guessed the old folder from the code's structure (Haiku 13, Sonnet 7 of 40). Deleting also kept the effort cost where the line was needed: on the test command, Sonnet used as many tool calls without the line as with the stale one (7.8 per run in both, against 5.8 with the corrected line; Table~\ref{tab:stage2}). Stale paths are also cheap to find: a script can check whether a path exists, although deciding whether a missing path is an error still needs judgment. We did not measure the effort of either repair.

\noindent\textbf{For agent and tool builders: the stale command failed silently.} Pytest found no tests and exited with an error, but the agents piped its output through \texttt{tail} or \texttt{head}, so the shell reported success. In 19 of Sonnet's 40 stale runs, the first check of the fix ran no tests, and Haiku often hid the error itself with \texttt{2>/dev/null} (16 of its 38 runs). Four safeguards follow, none of which we evaluated. A harness that loads an instruction file can flag paths in it that do not exist. A shell can set \texttt{pipefail}, so that a failing command piped into \texttt{tail} still fails, although this does not help when the agent hides the error. An agent can treat ``no tests ran'' as a failure. And an agent that works around a missing path can say so in its final summary and propose the one-line fix. Unlike a stale convention, a stale path can be detected without deciding whether the file or the code is right.

\noindent\textbf{For researchers: measure actions, and compare against both repairs.} Pass rates would have scored all three conditions the same. In four Command tasks, all 15 of Sonnet's runs even produced the same patch, byte for byte, so inspecting the patch would not help either. A trajectory measure that counts any test call as validation would score the stale command, which ran no tests, as a verified fix. Studies of instruction files should therefore report what agents did, such as operations on paths that no longer exist, and compare against both a corrected and a removed version of the instruction. In our study, stale references changed what agents did but not whether they fixed the bug, so evaluations that score only the delivered code would miss this effect.

\section{Threats to Validity}\label{sec:threats}
We list the main threats to our conclusions and how we addressed them.

\noindent\textbf{Number of cases.} We chose depth over breadth: each case is a real-world stale line with a documented history, studied under three conditions with the code and tasks held fixed. Because instruction type, repository, and tasks vary together, their separate effects cannot be told apart.

\noindent\textbf{Task difficulty.} Most tasks are small seeded defects in the code each stale line describes, admitted only if the repository's own tests detect them. Harder maintenance work may show different costs.

\noindent\textbf{Statistical inference.} The \exRuns{} runs repeat three cases rather than sampling independent instances of debt, and per-task estimates are imprecise. We therefore use exact tests stratified by task, correct the primary tests for multiplicity, treat an undetected difference as unresolved rather than absent, and label analyses added after the protocol was fixed as exploratory. While checking our numbers, we found that the stale-path detector missed relative paths after a change of directory; we corrected it, which changed 19 Sonnet runs in the Procedure case, and the replication package contains the results before and after the fix.

\noindent\textbf{Generalization.} This is a focused experiment: two models from one vendor in one harness, three cases of stale context, and \exTasks{} tasks. It shows how stale instructions acted in these settings, but other models, harnesses, kinds of stale context, and tasks may behave differently, so our findings should be read as evidence about these cases rather than as general rates.

\section{Conclusion}\label{sec:conclusion}

A stale line in an agent instruction file is documentation debt whose interest is paid in behavior rather than in outcomes. Across three real-world stale instructions, two models, and \exRuns{} runs, a stale test command and a stale procedure sent agents to paths that no longer exist, while a stale code location, which no task needed, had no effect. Agents still fixed the bug in nearly every run, so task success hid the debt: the stale test command checked nothing yet appeared to succeed, and agents worked around the stale path without repairing the file. Correcting the line removed this wasted work, and deleting it removed most of it. Future work should test these findings with more stale instructions, repositories, languages, models, and agent harnesses, measure how often agents reach stale lines in ordinary work, and evaluate the safeguards we propose. Until then, evaluations of instruction files should report what agents do, not only whether they succeed.

\section*{Data Availability}
All data and code of this study are publicly available through our replication package~\cite{replication2026}.

\balance
\bibliographystyle{IEEEtran}
\bibliography{references}
\end{document}